\documentclass[usenatbib,usegraphicx]{mn2e}
\usepackage{multirow}
\usepackage{booktabs}
\def\doh{$\Delta\rmn{log(O/H)}$}
\def\hi{H{\,\sc i}}
\def\hr{H{\,\sc ii} region}
\def\hrs{H{\,\sc ii} regions}
\def\ms{\,M$_\odot$}
\title[Predictions of the extent of self-enrichment in oxygen of giant metal-poor H\,\it{\small II} regions.]{Predictions of the extent of self-enrichment in oxygen of giant metal-poor H\,{\sc ii} regions.}
\author[Aida Wofford]{Aida Wofford$^{1}$\thanks{E-mail:nava@nhn.ou.edu}\\
$^{1}$Homer L. Dodge Department of Physics and Astronomy\\The University of Oklahoma, 440 West Brooks, Room 131, Norman OK, 73019-2061, USA}
\begin{document}
\date{Accepted 2008 January 15. Received 2008 November 1; in original form 2008 May 12}
\pagerange{\pageref{firstpage}--\pageref{lastpage}} \pubyear{2008}
\maketitle
\label{firstpage}
\begin{abstract}
In general, \hrs~do not show clear signs of self-enrichment in products from massive stars ($M \ge 8$\ms). In order to explore why, I modeled the contamination with Wolf-Rayet (WR) star ejecta of metal-poor ($Z=0.001$) \hrs, ionised either by a 10$^6$-M$_\odot$ cluster of coeval stars (cluster 1), or a cluster resulting from continuous star formation at a rate of 1\ms\,yr$^{-1}$ (cluster 2). The clusters have $Z=0.001$ and a Salpeter initial mass function (IMF) from 0.1 to 120\ms. Independent one dimensional constant density simulations of the emission-line spectra of unenriched \hrs~were computed at the discrete ages 1, 2, 3, 4, and 5\,Myr, with the photoionisation code CLOUDY \citep{fer98}, using as input, radiative and mechanical stellar feedbacks predicted by the evolutionary synthesis code STARBURST99 \citep{lei99,vaz05}. Each \hr~was placed at the outer radius of the adiabatically expanding superbubble of \cite{mac88}. For models with thermal and ionisation balance time-scales of less than 1\,Myr, and with oxygen emission-line ratios in agreement with observations, the volume of the superbubble and the \hr~were uniformly and instantaneously polluted with stellar ejecta predicted by STARBURST99. I obtained a maximum oxygen abundance enhancement of 0.025\,dex, with cluster 1, at 4\,Myr. It would be unobservable. 
\end{abstract}
\section{INTRODUCTION}\label{sec:int}
Single massive stars, which have masses $M \ge 8$\ms~\citep[\S~2.1.3]{mat01}, play a seminal role in the chemical enrichment of galaxies and the intergalactic medium. Indeed, they produce the bulk of the $\alpha$-elements (O, Ne, Mg, Si, S, Ar, Ca, and Ti) and they contribute to some Fe and Fe-peak elements \citep[\S~4]{fra04}. In addition, at solar  metallicity\footnote{Metallicity, $Z$, is the mass fraction of metals, i.e., of elements heavier than He. Throughout this work, and for consistency with the STARBURST99 models, the reference solar metallicity is $Z_\odot\sim0.020$ \citep{and89}. Since oxygen is the most abundant metal in the universe, the number density ratio O/H is often used as a gauge of metallicity. For $Z_\odot\sim0.020$, $12+\rmn{log[O/H]_\odot}\sim8.90$. Note that a more recent value is $12+\rmn{log[O/H]_\odot}=8.66\pm0.05$ \citep{asp04}.}, most of them synthesize all elements from B to Rb \citep{lim07}. Finally, at very low metallicity, i.e., at $12+\mathrm{log(O/H)}<7$, they produce significant amounts of C and N (see \citealt{pet08} for C; see \citealt{chi06} for N), although at the rest of metallicities, their contribution to the latter two elements, which are essential to life, is still under investigation (see \citealt{car05} versus \citealt{mat06} for C; see \citealt{hen06} for N).

Unfortunately, the chemical yields of such stars are rather uncertain (e.g., \citealt{mey94}). This issue should be resolved as we learn more about stellar mass loss from theory and observations (e.g., as in \citealt{ful06}), and as we improve our understanding of how specific burning reaction rates affect nucleosynthesis (e.g., \citealt{tur07}). In addition, observations of environments which have become polluted with massive star ejecta, on space-scales of a few hundred parsecs, and over relatively small time-scales ($t\le 10^7\,\rmn{yr}$), i.e., of self-enriched environments, should contribute to calibrate massive star yields. 

Targets associated with massive stars which could be observed in order to accomplish this task are: the circumstellar nebulae produced by the winds of red supergiants (RSGs) or luminous blue variables (LBVs), the circumstellar bubbles formed by the winds of WR stars, supernova remnants (SNRs), the superbubbles formed by the winds and supernovae (SNe) of massive stars in associations, and \hrs, in particular, low metallicity \hrs, since self-enrichment is expected to be more noticeable when the content in metals of the `to be contaminated' gas is initially low. 

Although our focus is on \hrs, which are among the most reliable targets for obtaining Galactic and extra-galactic chemical abundances, the rest of the above objects may affect the spectra of giant and extragalactic \hrs. For this reason, let us briefly review enrichment in all of them.

After the main sequence (MS), massive stars become RSGs, or if they have very large birth masses ($M\sim60\,\rmn{M}_\odot$), then they become LBVs \citep{not97}. During these stages, stars turn on slow but dense winds. These winds slowly expand away, producing circumstellar nebulae enriched with processed stellar material \citep{gs96a, gs96b, smi98, geh04, hir07}. Later on, if the star's birth mass is $M \ge 30\,\rmn{M}_\odot$ at solar metallicity \citep{lim07}, or higher at a lower metallicity, then the star will become a WR star. During this stage, the star turns on an enriched wind again. This fast wind sweeps up the previous circumstellar nebula, which collects in a circumstellar shell known as a wind bubble \citep{gs96a, gs96b, chu04, art07}.

Moving to mass loss at the end of the stellar lifetime, X-ray observations can show enrichment in individual young SNRs. \cite{rak06} used radio and X-ray observations of the ejecta-rich Galactic SNR G337.2-0.7 in order to determine that the origin of the remnant is a SNIa and not a core collapse SN. In the optical, although abundance studies of SNRs and \hrs~in the spiral galaxies M33 and M31 by \cite{dop80} and \cite{bla82}, respectively, revealed no signs of enrichment in N at a given radius, recent optical observations of SNRs show N-enrichment \citep{rud07}, as well as O-enrichment \citep{gha05} from core collapse SNe. 

Finally, when multiple stars of spectral type O and B are gravitationally bound in a cluster or association, then their individual circumstellar shells expand inside the superbubble blown collectively by their winds and SNe \citep{mac88,chu04,cho08}. The interiors of such superbubbles are filled with hot ($10^6-10^7\,\rmn{K}$) gas whose origin is shocked fast wind which then mixes in with SN ejecta. \cite{sil01} predicted that superbubble enrichment should be observable within the first $10^7\,\rmn{yr}$ of the superbubble's lifetime. However, no convincing cases of self-enrichment have been reported from X-ray observations of the hot gas inside superbubbles in our Galaxy or in the Magellanic Clouds. One plausible explanation is that some of the interstellar gas swept up by the superbubble, which has a temperature of $T\le10^4\,\rmn{K}$, i.e., which is cooler than the million-degree gas inside the superbubble, evaporates into the hot superbubble's interior, diluting its enriched content. 

We now return to \hrs. These ionised nebulae can be found around circumstellar bubbles and superbubbles \citep{tow03,chu04}. First we review some of the evidence showing that these objects do not become significantly polluted with massive star ejecta during their lifetime ($t \sim 10^7$ yr). Measurements in the Galactic disk, as well as in other spirals, indicate that fluctuations in the metal abundances of \hrs~located at the same galactic radius are less than $20-30$ per cent on average \citep{sca04}. For instance, based on a study of 41 \hrs~in M101, \cite{ken96} suggest that the dispersion in the $\mathrm{log(O/H)}$ abundance is considerably less than $\pm 0.1- \pm 0.2\,\rmn{dex}$. In addition, localized enrichment does not occur in the Orion nebula, which is only a few hundred parsecs away from us. Indeed, \cite{est04} measured heavy element abundances in this \hr~and found that they are just slightly above solar. Regarding low metallicity \hrs, a study by \cite{kob99} of nearly 100 low-mass, metal-poor starburst galaxies, aimed at detecting He-, N- and O-pollution by WR stars, in \hrs, revealed no significant variations over spatial scales $l>10\,\rmn{pc}$. \cite{kob99} offered the following three explanations for the lack of success in the study: (1) an IMF abnormally truncated at the high-mass end, or steeper, which would reduce the amount of contaminating ejecta from massive stars, (2) stellar black holes, which would swallow the polluting ejecta, (3) ejecta predominantly located in the cold phase ($T=10^2\,\rmn{K}$), or alternatively, in the hot phase ($T=10^6\,\rmn{K}$) of the interstellar medium (ISM). 

\cite{kob99} rejected the idea of an `abnormal' IMF, on the basis of the undeniable presence of massive WR stars. Stellar black holes on the other hand are always a possibility. It would be interesting if they can be detected. Finally, the hypothesis that the ejecta is predominantly located in the hot phase of the ISM, is the basis of the model of \cite{ten96}. Indeed, in the latter model, the ejecta from OB stars in associations remains trapped within the supperbubble carved in the ISM by stellar winds and type II supernova (SNII) explosions, while the superbubble expands through the ISM. This is because two gases can only mix through diffusion efficiently if they have similar high temperatures and similar densities, but the conditions in the superbubble interior ($T=10^6-10^7\,\rmn{K}$, $n=10^{-3}-10^{-2}\,\rmn{cm}^{-3}$) and in the immediate ISM, i.e., in the ISM swept-up by the superbubble ($T\le10^4$ K, $n=10^0-10^2$ cm$^{-3}$), are different. The ejecta is only released by the superbubble when the superbubble reaches the halo of its host galaxy and breaks up due to the negative density gradient. The ejecta may then blow out of the galaxy or it may rain back in the form of cold `droplets' of $10^{13}-10^{15}\,\rmn{cm}$ in size \citep{sta07}, where molecules and grains have formed. These droplets spread over kiloparsec scales and mix at the atomic level with the ISM when the radiation from a new generation of massive stars dissociates the droplet molecules and grains, 10$^8$\,yr after the matter was ejected from the stars. Note that interstellar metal enhancements, such as the droplets, would be hard to observe, given their size, in particular if significant dilution of the stellar ejecta with interstellar material occurs within the superbubble.    

However, there is some evidence that \hrs~become polluted with massive star ejecta, during their lifetimes, at most metallicities. \cite{kob96} found a fluctuation of $0.095\pm0.019$ dex in $\mathrm{log(O/H)}$ over a scale of 200 pc that is consistent with pollution from SNeII, between two starburst regions of NGC 4214. \cite{sta03} detected possible enrichments in N and maybe in O in a study of 400 H II galaxies spanning the metallicity range $Z/Z_\odot=0.02-0.2$. The oxygen abundances recently determined within an area of $400\times300\,\rmn{pc}^2$, in the H II galaxy II Zw 70, span the range $12 + \mathrm{log(O/H)} = 7.65-8.05$ \citep{keh08}, and the contribution of pollution with massive star ejecta to this scatter is still to be determined. Also, \cite{tsa05} have argued that 30 Doradus in the LMC, which is the closest giant \hr~to us, is affected by pollution from SN ejecta. Furthermore, a recent model of a wind and radiation driven bubble around a solar metallicity star of $85\,\rmn{M}_\odot$, shows that at the end of the stellar lifetime, the amount of C reaches 22.3 per cent above solar in the associated \hr, although the amounts of O and N, in the same location, are `insignificantly higher than solar' \citep{kro06}. Finally, the most well established case of \hr~self-enrichment is that of the starburst galaxy NGC 5253, where N was reported to be larger by a factor of 3 at two locations with respect to a third one, within a length-scale of $50-60$ pc \citep{kob97}. This enrichment, which seems to be accompanied by an enhancement in He, in agreement with massive star evolution, was recently re-established by \cite{lop07}, and is probably associated with the circumstellar environment of underlying WR stars. Note that galaxy Markarian 996 could be a case similar to that of NGC 5253 \citep{thu96}. In summary, the extent to which \hrs~become self-enriched during their lifetime is still under investigation.

As can be seen from the above summary, predictions of the physical and spectral properties of self-enriched \hrs, which are useful for learning more about the fate of massive star products, can only be as good as our understanding of how such stars evolve and affect their circumstellar environment. In order to move forward in our understanding of the chemical enrichment of the ISM, I predicted oxygen abundance enhancements, due to pollution with massive star ejecta, of metal-poor ($Z=0.001$) \hrs, with ages $t\le5$\,Myr. The reason for studying oxygen is that this element is the dominant metal in massive star ejecta and a dominant coolant in \hrs, therefore, it produces strong observable spectral lines. The reason for choosing $Z=0.001$ is that self-enrichment is expected to be more noticeable when the initial content in metals of the ISM is low. The reason for selecting ages less than 5\,Myr is that I concentrated on pollution with WR star ejecta. I considered clouds of different gas densitites and ionised by star clusters resulting from different star formation laws (SFLs). I analyzed the following steady-state properties of these clouds: (a) time-scale to reach thermal and ionisation balance; (b) values of the [O\,{\sc i}]\,$\lambda6300$/[O\,{\sc ii}]\,$\lambda3727$ and [O\,{\sc ii}]\,$\lambda3727$/[O\,{\sc iii}]\,$\lambda\lambda4959,5007$ emission-line strength ratios, $R12$ and $R23$, respectively; (c) thickness of the \hr; and (d) contribution \doh~of massive star ejecta to the total oxygen abundance of the \hr. I studied how the evolution of the spectral energy distribution of the stars, the increase in the separation between the \hr~and the stars, and the dust content and hydrogen density of the \hr, affect properties (a) to (d). As a test of the validity of each model, the ratios $R12$ and $R23$ were compared with observed values collected by \cite{nav06}. The paper is organized as follows. In \S~\ref{sec:mod}, I outline the capabilities of the two computer codes used in this study and provide the input parameters and model assumptions. In \S~\ref{sec:res}, I discuss the output from the computer codes and explain my results for (a) to (d). In \S~\ref{sec:con}, I summarize and conclude. In order to make the paper easier to read, details on the superbubble model of \cite{mac88}, and part of the discussion on the model assumptions are provided in Appendices A and B. 
\section{Models}\label{sec:mod}
Here I outline the capabilities of the two computer codes used in my work and I provide the input parameters and model assumptions.
\subsection{STARBURST99}\label{mod:s99}
STARBURST99 \citep{lei99,vaz05}, is a widely used evolutionary synthesis code for modeling various properties of stellar populations. Initially, the code was optimized for young stellar populations ($t\le10^8\,$yr), which coined its name. However, the current running version (5.1) allows to model older stellar populations, as described in \cite{vaz05}. Different choices are available for the following input parameters: star formation law and related quantities, metallicity and stellar evolution tracks, and stellar atmospheres and spectra. I used version 5.1 of the code for predicting the radiative, mechanical, and chemical feedbacks of ionising stellar populations, between 0.1 and 5\,Myr. Here is my choice of input parameters.
\subsubsection{Star formation law}\label{s99:sfl}
I tested two star formation laws, which according to \cite{lei99}, reproduce well typical spectrophotometric properties of star-forming regions in galaxies, i.e., a single burst of star formation with a total birth mass of 10$^6\,$M$_\odot$, where the stars are born simultaneously and instantaneously (hereafter, instantaneous SFL), and continuous star formation at the constant rate of 1\ms~yr$^{-1}$ (hereafter, continuous SFL). For instance, a 10$^6\,$\ms-cluster is thought to be responsible for an ultra-compact \hr~in starburst galaxy NGC 5253 \citep{tur04}. Henceforth, the stellar populations resulting from the instantaneous and continuous SFLs will be referred to as clusters 1 and 2, respectively. 
\subsubsection{Initial mass function}\label{s99:imf}
Stars span the mass range from 0.1 to 120\ms~and are distributed according to a Salpeter IMF \citep{sal55}, i.e., such that the initial number of stars per unit volume (in pc$^3$) and with masses between $m$ and $m+dm$ (in M$_\odot$) is given by $dN = k~m^{-2.35}~dm$, where the constant $k$, normalizes the mass distribution to the total stellar mass. Given this IMF, the initial number of O V stars in cluster 1 is $\sim$2$\times10^3$.
\subsubsection{Final stages of stellar evolution}\label{s99:com}
The minimum mass of a SNII progenitor was set to 8\ms, and the cut-off mass for the formation of black hole to 120\ms, i.e., no black holes are formed. The choice of no black holes maximizes the total amount of contaminating material for the \hr, since the ejecta from supernovae are not swallowed by the remnants.
\subsubsection{Metallicity and tracks}\label{s99:z}
I used the set of evolutionary tracks best suited for young stellar populations, i.e., the set of non-rotating Geneva tracks. Since self-enrichment is expected to be more noticeable at low metallicity, I selected the lowest metallicity available for these tracks, i.e., $Z=0.001$. The adopted set of tracks uses the standard mass loss models of \cite{sch92} for stars with masses $M<12\,$M$_\odot$, and the high mass loss models of \cite{mey94} for stars with $M\ge12\,$M$_\odot$. The tracks of \cite{mey94} cover all phases up to the end of central C burning and reproduce well most of the observed properties of WR stars, in particular, the WR subtype distributions and the low luminosities reached by certain WR stars, although they do not reproduce well the mass loss rate itself \citep{lei95}.
\subsubsection{Chemical yields}\label{s99:chem}
The wind chemical yields, i.e., the mass in each element ejected by winds in \ms~yr$^{-1}$, come straight out of the evolutionary tracks. Note that since only H, He, C, N, and O have significant changes in the pre-supernova phase, the wind yields for elements like Si or Fe are simply determined by the initial abundances. On the other hand, the yields from SNeII are those of \cite{ww95}, with the corrections of \cite{lin99}.
\subsubsection{Stellar atmospheres and spectra}\label{s99:atm}
In STARBURST99, the ensemble of stars is treated as a single source with a luminosity equal to the sum of the luminosities of the individual stars, and with a spectral shape equal to the superposition of the spectra of the stars. The individual stellar spectra come from \cite{lej97} for stars with plane-parallel atmospheres, \cite{pau01} for O stars, and \cite{hil98} for other stars with strong winds. The cluster spectra span the wavelength range $91-1.6\times10^6\,\rmn{\AA}$. They are composed of $1220$ points, which are spaced such that $\Delta\lambda=2-10\,\rmn{\AA}$ for $\lambda\le912\,\rmn{\AA}$ and $\Delta\lambda=10-2\times10^5\,\rmn{\AA}$ for $\lambda>912\,\rmn{\AA}$.
\subsection{CLOUDY}\label{mod:cl}
CLOUDY \citep{fer98}, is a widely used plasma/molecular photoionisation code that predicts the thermal, ionisation, and chemical structures of clouds, ranging from intergalactic to the high-density local-thermodynamic-equilibrium limit, and their spectra. The code requires the following input parameters: 1) shape and luminosity of the external radiation field striking the cloud (hereafter, the incident continuum), 2) chemical composition and grain content of the gas, and 3) geometry of the gas, including the radial extent of the gas and the dependence of its density on the radius. I computed one dimensional constant density steady-state models of the emission-line spectra of \hrs, with version 07.02 of CLOUDY, using for 1) the spectral energy distributions predicted by STARBURST99 at times $t=1$, 2, 3, 4, and 5\,Myr since the beginning of star formation, and for 3) the mechanical feedback of the clusters, as explained in \S~\ref{cl:geom}. Note that the \hr~models are independent of each other, i.e., the steady-state properties of a given cloud do not affect the properties of subsequent models. The computation of time-dependent \hr~simulations may be possible with the 2009 release of CLOUDY (G. Ferland, private communication). Further model details are provided below.
\begin{table*}
\centering
\caption{Gas and dust abundances in units of $12+\rmn{log}[n/n(\rmn{H})]$, and scaling factors applied to $n/n(\rmn{H})$.}
\begin{tabular}{l@{\hspace{0.25cm}}ccclccc}
\toprule
\multicolumn{1}{l}{element}&
\multicolumn{3}{c}{reference value}&
\multicolumn{2}{c}{scaling factor}&
\multicolumn{2}{c}{value of model cloud}\\
\cmidrule(r){2-4}\cmidrule(lr){5-6}\cmidrule(lr){7-8}
\hfill&
sun\footnote{Mean of photospheric and meteoritic values corresponding to $Z_\odot=0.019$ \citep{and89}.}&
dust\footnote{CLOUDY's default dust grain abundances. They include a graphitic and silicate component, exclude polycyclic aromatic hydrocarbons, and generally reproduce the overall extinction properties of the Galactic ISM for a ratio of extinction per reddening of $R_V=3.1$ \citep[][see Hazy 1, p.\,61]{haz06}.}&
dust/sun\footnote{Ratio of dust and solar abundances.}&
gas\footnote{The first factor depletes the reference solar abundance by 20, because the model metallicity is $Z\approx Z_\odot/20=0.001$. The second factor, if present, accounts for the depletion of the element from the gas, due to its presence in the dust. Note that although CLOUDY's dust does not include Na, Al, Ca, and Ni, these elements are expected to be depleted in the gas phase, and present in the dust. Also note that although CLOUDY's dust includes C and O, these elements were not depleted from the gas, in order to account for their presence in the dust.}&
dust\footnote{This factor depletes CLOUDY's default dust abundance by 20, because the model metallicity is $Z\approx0.001$.}&
gas&
dust\\
\toprule
H & 12.00 & - & - & - & - & 12.00 & - \\
He\footnote{The helium gas abundance was derived from Eqns. (\ref{eqn:Y}) and (\ref{eqn:y}).} & 10.99 & -& - & - & - & 10.90 & - \\
C & 8.56 & 8.44 & 0.76 & 0.05 & - & 7.26 & 7.14 \\
N & 8.05 & - & - & 0.05 & - & 6.75 & - \\
O & 8.93 & 8.12 & 0.15 & 0.05 &-  & 7.63 & 6.82 \\
Ne & 8.09 & - & - & 0.05 & - & 6.79 & - \\
Na & 6.32 & - & - & 0.05$\times$0.1 & - & 4.02 & - \\
Mg & 7.58 & 7.52 & 0.87 & 0.05$\times$0.1 & 0.05 & 5.28 & 6.21 \\
Al & 6.47 & - & - & 0.05$\times$0.1 & - & 4.17 & - \\
Si & 7.55 & 7.52 & 0.93 & 0.05$\times$0.1 & 0.05 & 5.25 & 6.21 \\
S & 7.24 & - & - & 0.05 & - & 5.94 & - \\
Ar & 6.56 & - & - & 0.05 & - & 5.26 & - \\
Ca & 6.35 & - & - & 0.05$\times$0.1 & - & 4.05 & - \\
Fe & 7.59 & 7.52 & 0.85 & 0.05$\times$0.1 & 0.05 & 5.29 & 6.21 \\
Ni & 6.25 & - & - & 0.05$\times$0.1 & - & 3.95 & - \\
\bottomrule
\label{tab:abun}
\end{tabular}
\end{table*}
\subsubsection{Gas and dust chemical compositions}\label{cl:abun}
The \hr~gas is composed of the 15 most abundant elements by number in the sun. All of these elements, except for Na, Al and Ca, have lines in the wavelength range $3,100-10,400\,\rmn{\AA}$ that have been observed in Galactic \hrs~\citep{g-r04}. On the other hand, Na, Al and Ca are refractory elements expected to be present in the form of dust in the ISM \citep{sem95,sem96, haz06}. The exclusion of additional elements should have little impact on the results, given the low metallicity of the clouds ($Z\approx0.001$). The abundances of elements in the gas and dust, used as input to the CLOUDY models, are given in columns 7 and 8 of Table~\ref{tab:abun}, respectively. They were computed as follows. The gas abundances were obtained by applying the scaling factors in column 5 to the solar values in column 2. Similarly, the dust abundances were obtained by applying the scaling factors in column 6 to the values in column 3, which are CLOUDY's default abundances for the Galactic ISM. Column 4 of Table~\ref{tab:abun} shows that C and O are present in CLOUDY's dust, and that dust has a high content of C. However, because dust abundances are highly uncertain, I did not deplete C and O from the gas in order to account for their presence in the dust, as I did for less abundant elements. In \S~\ref{sec:res}, I show that the effect of dust on the emitted spectrum of the \hrs~in negligible. This is due to the low metallicity of the models.

For consistency between CLOUDY and STARBURST99, the helium mass fraction $Y$ was computed from
\begin{equation}
Y=Y_P+\left(\frac{Y_\odot-Y_P}{Z_\odot-Z_P}\right)Z,\label{eqn:Y}
\end{equation}
where the primordial values are $Y_P=0.24$ and $Z_P=0$, while the solar values are $Y_\odot=0.299$ and $Z=0.019$ \citep{sch92}. Equation (\ref{eqn:Y}) yields $Y=0.243$. The ratio of the helium to hydrogen number abundances, $y=n(\rmn{He})/n(\rmn{H})$, was derived from Eqn. (\ref{eqn:Y}) and
\begin{equation}
Y=\frac{M(\rmn{He})}{M(\rmn{H})+M(\rmn{He})+M(Z)}\approx\frac{4\,y}{1-4y},\label{eqn:y}
\end{equation}
where $M$(H), $M$(He), and $M(Z)$ are the masses in H, He and metals, and the right hand side of the Eqn. (\ref{eqn:y}) comes from neglecting $M(Z)$ and using $M(\rmn{H})=m(\rmn{H})\,n(\rmn{H})\,V$ and $M(\rmn{He})=4\,m(\rmn{H})\,n(\rmn{He})\,V$, where $m(\rmn{H})$ is the atomic mass of H, and $V$ is some volume. The resulting He abundance is $n(\rmn{He})\approx0.1\,n(\rmn{H})$.
\begin{figure}
\begin{center}
\includegraphics[width=0.3\textwidth]{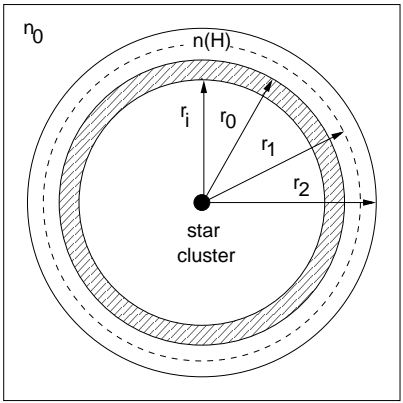}
\caption{Model geometry. Hashed shell: bounding shell of superbubble. $r_i$: inner radius of superbubble's shell. $r_0$: outer radius of superbubble's shell. Region between $r_0$ and $r_2$: model cloud. Region between $r_0$ and $r_1$: \hr. Dashed-circle: \hr's ionisation front.\label{fig:geom}}
\end{center}
\end{figure}
\subsubsection{Geometry}\label{cl:geom}
The ionising cluster is located at the center. The cloud has a covering factor of $\sim$4$\pi$ (i.e., the central source is fully covered), a filling factor of one (i.e., there are no clumps in the gas), a constant value of $n$(H) (cm$^{-3}$), and it starts at distance $r_0$ from the center. The value of $r_0$ increases with time as given by the model of \cite{mac88} for the adiabatic expansion of a superbubble [see Appendix~A for a summary]. The outer radius of the superbubble, $r_0$, is given by
\begin{equation}
r_0=(267\,\rmn{pc})\left(\frac{L_{38}\,t_7^3}{n_0}\right)^{1/5},\label{eqn:r0}
\end{equation}
where $L_{38}$ ($10^{38}$ erg s$^{-1}$) is the average mechanical luminosity from the stellar winds and supernovae (I used the average from 0.1 to 5\,Myr), $t_7$ ($10^7$ yr) is the time since the onset of stellar mass loss (it starts on the main sequence for massive stars), and $n_0$ (cm$^{-3}$) is the density of the ambient undisturbed ISM at the location of the superbubble (I computed models with $n_0=1$ and $n_0=10$). Note that by letting the \hr~start at $r_0$, I am assuming that its ionisation front is located outside of the superbubble. In nature, the ionisation front may be located within the superbubble. In \S~\ref{sec:res}, I compare the value of $r_0$ with the Str\"omgren radius of each \hr.

The radial extent of the cloud is set by the point where the electron temperature of the gas has dropped to 100 K. For most models, the resulting geometry of the emission is plane-parallel. Figure~\ref{fig:geom} is a sketch of the gas geometry. As shown in the figure, the thickness of the superbubble's bounding shell is $r_0-r_i$, the thickness of the model cloud is $r_2-r_0$, and the thickness of the \hr~is $r_1-r_0$.
\subsubsection{Volume of radius $r_0$}\label{cl:r0}
In the superbubble model of \cite{mac88}, the volume of radius $r_i$ contains gas with a temperature $T$ given by
\begin{equation}
T=T_c(1-r/r_0)^{2/5},\label{eqn:t}
\end{equation}
where
\begin{equation}
T_c=(3.5\times10^6\,\rmn{K})L_{38}^{8/35}\,n_0^{2/35}\,t_7^{-6/35}\label{eqn:tc}
\end{equation}
is the central temperature; and an atomic density $n$ given by
\begin{equation}
n=n_c(1-r/r_0)^{-2/5},\label{eqn:n}
\end{equation}
where
\begin{equation}
n_c=(4\times10^{-3}\,\rmn{cm^{-3}})L_{38}^{6/35}\,n_0^{19/35}\,t_7^{-22/35}\label{eqn:nc}
\end{equation}
is the central density. I assumed that the value of $n$ at $r_i$ is constant for $r_i\le r\le r_2$, and I used Eqns. (\ref{eqn:t}) to (\ref{eqn:nc}) to derive the hydrogen density $n$(H) of the model cloud, as explained next. Note that for the purpose of radiative transfer, the volume of radius $r_0$, $V_0$, was treated as a cavity empty of gas. This assumption is discussed in Appendix~B. 
\subsubsection{Hydrogen density}\label{cl:nh}
For simplicity, I assumed that $n$(H) is constant throughout the model cloud. However, I assumed that $n$(H) varies with time as given by
\begin{equation}
n(\rmn{H})=\frac{1.4\times L_{38}^{2/5}\,n_0^{3/5}}{1.1\times t_7^{4/5}}.\label{eqn:nh}
\end{equation}
I obtained Eqn. (\ref{eqn:nh}) as follows. First I solved Eqn. (\ref{eqn:t}) for the radius, which yields, $r=r_0\left[1-(T/T_c)^{5/2}\right]$. Then I substituted the latter expression in Eqn. (\ref{eqn:n}), which yields $n=n_c\,T_c/T$. Next, I used $n\approx n(\rmn{H})+n(\rmn{He})\approx1.1\times n(\rmn{H})$. Finally, I assumed that $T=10^4\,$K, which is the characteristic equilibrium temperature of an \hr. In summary, the atomic density of the model cloud was assumed to be equal to the atomic density given by Eqn. (\ref{eqn:n}), at the point where $T=10^4\,$K. 
\subsection{Self-enrichment}\label{cl:enrich}
Let $\rmn{O/H}\equiv n(\rmn{O})/n(\rmn{H}) $. In addition, let $i$ and $f$ stand for `initial' and `final', where initial is `prior to' and final is `after' contamination of the cloud with stellar ejecta. Finally, let $\Delta n$ be the increase in the number density of an element in the \hr, due to contamination with massive star ejecta. The contribution \doh~of the ejecta to the oxygen abundance of the \hr, is given by 
\begin{equation}
\Delta\rmn{log}\frac{\rmn{O}}{\rmn{H}}=\rmn{log}\left(\frac{\rmn{O}}{\rmn{H}}\right)_f-\rmn{log}\left(\frac{\rmn{O}}{\rmn{H}}\right)_i=\rmn{log}\left(\frac{\rmn{O}}{\rmn{H}}\right)_f+4.371,
\end{equation}
where
\begin{equation}
\left(\frac{\rmn{O}}{\rmn{H}}\right)_f=\frac{n_i(\rmn{O})+\Delta n(\rmn{O})}{n_i(\rmn{H})+\Delta n(\rmn{H})}.
\end{equation}
The value of $n_i(\rmn{H})$ is given by Eqn. (\ref{eqn:nh}), that of $n_i$(O) by column 7 of Table~\ref{tab:abun}, and that of $\Delta n(\rmn{X})$ by
\begin{equation}
\Delta n(\rmn{X})=\frac{M(\rmn{X})}{(V_0+V_1)\,m(\rmn{X})},\label{eqn:deln}
\end{equation}
where $M(\rmn{X})$ is the cumulative mass in element X ejected by the stars, up to the time under consideration, $V_0$ is the volume of radius $r_0$, $V_1$ is the volume of the shell of thickness $r_1-r_0$, and $m(\rmn{X})$ is the atomic mass of element X.

At this point, one may wonder what is the physical mechanism assumed for the mixing of the stellar ejecta with the \hr. Turbulent mixing is expected to occur in shock fronts and contact discontinuities, where gas is stretched and folded until the gradient length approaches the collision mean free path \citep{sca04}. I assumed that stellar winds and supernovae disperse the ejecta throughout the volume of radius $r_i$, and that turbulence in the wind and SN shocks mixes the ejecta with the gas pre-existing in the volume, quickly and efficiently. On the other hand, according to \cite{ten96}, mixing across the superbubble's shell is inefficient. This is because two gases can only mix through diffusion efficiently if they have similar high temperatures and similar densities, but the conditions of the gas for $r\le r_i$ ($T\sim10^6-10^7$ K, $n\sim10^{-3}-10^{-2}$ cm$^{-3}$) and in the \hr~($T\approx10^4$ K, $n=10-10^2$\,cm$^{-3}$) are different. Without providing a mechanism for the mixing across the superbubble's shell, I test if uniform and instantaneous mixing of the stellar ejecta, with the gas in the volume $V_0+V_1$, results in an observable enhancement in the O/H abundance of the \hr.
\subsubsection{Supplementary models}
For each cluster and value of $n_0$, three supplementary models were computed at 4\,Myr: one without dust, one where the hydrogen density was arbitrarily set to $n(\rmn{H})=100$\,cm$^{-3}$, and one where the distance to the cloud was arbitrarily set to $r_0=100$ pc. The purpose of these models is to test the sensitivity of \doh~to the dust content of the cloud, $n(\rmn{H})$, and $r_0$.
\section{Results}\label{sec:res}
Here I discuss properties of the radiative, mechanical, and chemical feedbacks of clusters 1 (instantaneous SFL) and 2 (continuous SFL). Hereafter, superindexes 1 and 2 are used for quantities related to each cluster, respectively. In this section, I also describe the output from CLOUDY, compare it with observations, and discuss the contribution \doh~of massive star ejecta to the total oxygen abundance of the \hrs. 
\begin{figure}
\begin{center}
\includegraphics[width=0.4\textwidth]{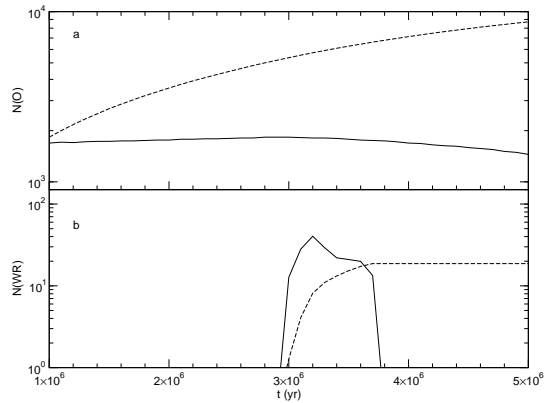}
\caption{Number of O stars [$N$(O), panel a] and WR stars [$N$(WR), panel b] versus time ($t$). Curves: solid, cluster 1; dashed, cluster 2. Time-step: 0.1 Myr.}\label{fig:nstars}
\end{center}
\end{figure}
\subsection{Number of O and WR stars}\label{res:nstars}
Figure~\ref{fig:nstars} shows evolutionary plots of the number of O stars [$N$(O), panel a] and WR stars [$N$(WR), panel b], for clusters 1 (solid curve) and 2 (dashed curve). As time increases, $N(\rmn{O})^1$ decreases while $N(\rmn{O})^2$ increases. Also, $N\rmn{(WR)^1}\ne0$ from 3 to 3.7 Myr, and $N\rmn{(WR)^2}\ne0$ for $t\ge3$\,Myr. Type O and WR stars affect the radiative, mechanical and chemical feedbacks of the clusters, as described below.
\begin{figure}
\begin{center}
\includegraphics[width=0.4\textwidth]{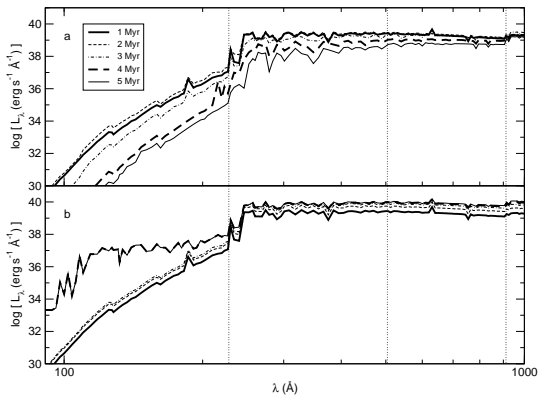}
\caption{Monochromatic luminosity ($L_\lambda$) versus wavelength ($\lambda$). Panels: a, cluster 1; b: cluster 2. Curves: thick-solid, 1 Myr; dashed, 2 Myr; dotted-dashed, 3 Myr; thick-dashed, 4 Myr; thin-solid, 5 Myr. Vertical-dotted lines at 228, 504, and 912 {\AA}, approximate ground-state ionising wavelengths of He{$\,$\sc ii}, He{$\,$\sc i} and H{$\,$\sc i}, respectively.}\label{fig:rad}
\end{center}
\end{figure}
\begin{table*}
\centering
\caption{Radiative, mechanical, and chemical feedbacks of clusters 1 and 2, at ages 1--5 Myr, predicted by STARBURST99. Entries (all logarithmic values except for time): time since the first stars formed, $t$; number of ionising photons in the H$\,${\sc i}, He$\,${\sc i}, and He$\,${\sc ii} continua, $Q$(H$\,${\sc i}), $Q$(He$\,${\sc i}), and $Q$(He$\,${\sc ii}); bolometric luminosity, $L_\rmn{bol}$; mechanical luminosity from winds and supernovae, $L_\rmn{mech}$; and masses in hydrogen and oxygen ejected by the stars up to time $t$, $M$(H) and $M$(O).}
\begin{tabular*}{1\textwidth}{@{\extracolsep{\fill}}*{8}{c}}
\toprule
t & 
Q(H$\,${\sc i}) & 
Q(He$\,${\sc i}) & 
Q(He$\,${\sc ii}) & 
L$_\rmn{bol}$ & 
L$_\rmn{mech}$ & 
M(H) & 
M(O) \\
Myr & 
s$^{-1}$ & 
s$^{-1}$ & 
s$^{-1}$ & 
erg s$^{-1}$ & 
erg s$^{-1}$ & 
M$_\odot$ & 
M$_\odot$ \\
\cmidrule(r){1-8}
\multicolumn{8}{c}{cluster 1} \\
\cmidrule(r){1-8}
1 & 52.563 & 52.034 & 48.321 & 42.398 & 38.467 & 1.934 & -1.222 \\
2 & 52.621 & 52.093 & 48.570 & 42.454 & 38.980 & 2.505 & -0.638 \\
3 & 52.499 & 51.843 & 47.907 & 42.527 & 39.392 & 3.301 & -0.009 \\
4 & 52.224 & 51.463 & 47.368 & 42.364 & 39.983 & 3.732 &  3.462 \\
5 & 51.959 & 51.056 & 46.224 & 42.203 & 39.989 & 3.833 &  3.462 \\
\cmidrule(r){1-8}
\multicolumn{8}{c}{cluster 2} \\
\cmidrule(r){1-8}
1 & 52.579 & 52.051 & 48.341 & 42.415 & 38.375 & 1.602 & -1.553 \\
2 & 52.888 & 52.359 & 48.672 & 42.723 & 38.900 & 2.342 & -0.824 \\
3 & 53.061 & 52.511 & 48.811 & 42.923 & 39.397 & 2.987 & -0.187 \\
4 & 53.142 & 52.574 & 49.494 & 43.050 & 40.390 & 3.672 &  3.176 \\
5 & 53.179 & 52.595 & 49.495 & 43.117 & 40.537 & 4.041 &  3.643 \\
\bottomrule
\end{tabular*}\label{tab:feedback}
\end{table*}
\subsection{Stellar radiative feedback}\label{res:rad}
Figure~\ref{fig:rad} shows plots of the monochromatic luminosity ($L_\lambda$) versus wavelength ($\lambda$), for clusters 1 (panel a) and 2 (panel b), at the discrete times 1, 2, 3, 4, and 5\,Myr, and in the region below the Lyman limit ($\lambda\le912\,${\AA}), which is the stellar spectral region of interest, since in \hrs, most of the H{\sc i} is the ground state, and therefore, the number of hydrogen ionisations by starlight photons with $\lambda>912\,${\AA} is negligible \citep{ost06}. Let $L_i$ be the ionising luminosity below the Lyman limit. The luminosity $L_i^1$ is dominated by the most massive O stars during the first Myr, by WR stars from 3 to 3.7\,Myr, and by less massive O stars from 3.7 to 5\,Myr. The value of $L_i^1$ decreases with time as the stars contributing to it fade away. On the other hand, $L_i^2$ increases with time for the first Myr, and then becomes independent of time, when a balance is established between the birth and death of the stars contributing to it (this is why in panel b of Fig.~\ref{fig:rad} the thick-dashed and thin-solid curves, which correspond to 4 and 5\,Myr respectively, overlap). Finally, $L_i^1<L_i^2$ from 1 to 5\,Myr. This is due to the continuous formation of ionising stars in cluster 2. In particular, the He$\,${\sc ii} continuum is a lot more prominent for cluster 2, from 4 to 5 Myr, due to the strong radiation in that spectral region, from WR stars, which are absent at those times, in cluster 1. 

Table~\ref{tab:feedback} complements Figs.~\ref{fig:nstars} and~\ref{fig:rad}. The entries in the table are: time since the formation of the first stars ($t$), number of ionising photons in the H$\,${\sc i}, He$\,${\sc i}, and He$\,${\sc ii} continua, [$Q$(H$\,${\sc i}), $Q$(He$\,${\sc i}), and $Q$(He$\,${\sc ii}) respectively], bolometric luminosity ($L_\rmn{bol}$), mechanical luminosity from winds and SNe ($L_\rmn{mech}$), and mass in oxygen ejected by the cluster up to the time considered [$M$(O)]. The evolutions of $L_\rmn{mech}$ and $M$(O) are discussed next.
\begin{figure}
\begin{center}
\includegraphics[width=0.4\textwidth]{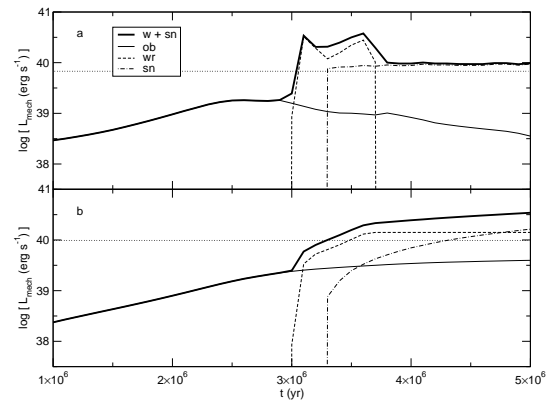}
\caption{Mechanical luminosity from stellar winds and SNe ($L_{mech}$) versus time ($t$). Panels: a, cluster 1; b, cluster 2. Curves: thick-solid, winds and SNe; thin-solid, OB stars; dashed, WR stars; dotted-dashed, SNe (Ib and II). Horizontal-dotted lines: average values $\rmn{log}\left<L_\rmn{mech}\right>$, from 0.1 to 5 Myr, for clusters 1 and 2. Time-step: 0.1 Myr.}\label{fig:mech}
\end{center}
\end{figure}
\subsection{Stellar mechanical feedback}\label{res:mech} 
Let $L_{mech}$ be the mechanical luminosity from stellar winds and SNe. Values of $L_{mech}^1$ and $L_{mech}^2$ at 1, 2, 3, 4, and 5\,Myr are given in columns 6 and 12 of Table~\ref{tab:feedback}, respectively. These columns show that $L_\rmn{mech}^1>L_\rmn{mech}^2$ at 1 and 2\,Myr, whereas $L_\rmn{mech}^1<L_\rmn{mech}^2$ at 3, 4, and 5\,Myr. Figure~\ref{fig:mech} shows plots of $L_\rmn{mech}$ versus time, for clusters 1 (panel a) and 2 (panel b). The contributions from different components have been plotted separately. The luminosity $L_\rmn{mech}^1$ is dominated by winds from OB stars from 1 to 3\,Myr, by winds from WR stars from 3 to 3.7\,Myr, and by SNe of type Ib and II from 4 to 5\,Myr. On the other hand, $L_\rmn{mech}^2$ is dominated by winds from OB stars from 1 to $~$3\,Myr, by winds from WR stars from $~$3 to $~$4.8\,Myr, and by SNe from $~$4.8 to 5\,Myr. The horizontal-dotted lines in Fig.~\ref{fig:mech} indicate the average values of $\rmn{log}\left<L_\rmn{mech}\right>$, from 0.1 to 5\,Myr, for clusters 1 and 2. Recall that the reason for adopting the average $\left<L_\rmn{mech}\right>$, is that Eqn. (\ref{eqn:r0}) for the radius $r_0$ is based on a constant input of mechanical energy from the stars. The values of $L_{38}=\left<L_\rmn{mech}\right>/10^{38}$ are $L_{38}^1\approx68$ and $L_{38}^2\approx97$. A comparison of the thick-solid curve and horizontal-dotted line in panel a of Fig.~\ref{fig:mech} shows that $L_{38}^1$ is overestimated for $t<3$\,Myr and underestimated for $t>3\,$Myr. The same is true for $L_{38}^2$, except that the limiting time is $t~3.2$\,Myr. This means that $r_0$ is underestimated and \doh~is overestimated at times when WR stars dominate the enrichment of the ISM. As shown in \S~\ref{res:doh}, that \doh~is overestimated at the latter times does not change the main conclusion of this study.
\begin{figure}
\begin{center}
\includegraphics[width=0.4\textwidth]{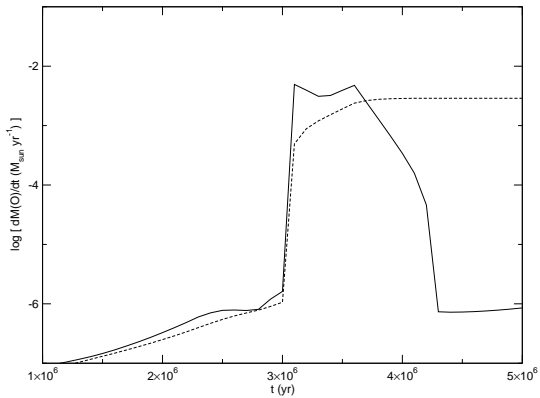}
\caption{Oxygen mass loss rate [$dM(\rmn{O})/dt$] versus time ($t$). Curves: solid, cluster 1; dashed, cluster 2. Time-step: 0.1 Myr.}\label{fig:chem}
\end{center}
\end{figure}
\subsection{Stellar chemical feedback}\label{res:chem}
Figure~\ref{fig:chem} shows plots of the oxygen mass loss rate $dM(\rmn{O})/dt$ (\ms\,yr$^{-1}$), for clusters 1 (solid curve) and 2 (dashed curve). The value of $[dM(\rmn{O})/dt]^1$ slowly increases from 1--3\,Myr when the winds from OB stars dominate the mass loss. Then $[dM(\rmn{O})/dt]^1$ rapidly increases when WR stars appear, and it maintains high values while these stars are present (from 3 to 3.7 Myr). Finally $[dM(\rmn{O})/dt]^1$ falls into a dip starting at $t\sim4.2\,$Myr, which is due to the fact that \cite{ww95} and \cite{lin99} did not publish SN yields for stellar masses higher than 40\ms, or equivalently, stellar ages below 5.6\,Myr. The justifications for the dip are that SNe with such high masses are very rare and do not contribute to the ISM enrichment; and also, that we do not know the yields of such events from observations. Note that in STARBURST99 v. 5.1, stars more massive than 40\ms~explode but without any nucleosynthesis, which explains why there is not a similar dip in the plot of $L_\rmn{mech}^1$. On the other hand, $[dM(\rmn{O})/dt]^2$ evolves similar to cluster 1 for the first Myr, but reaches a steady value around $t=4$\,Myr, which is maintained by the constant presence of WR stars.
\begin{table*}
\centering
\caption{Input and output of the CLOUDY models for the instantaneous and continuous SFLs (cluster 1 and 2, respectively), and for the two values of the atomic density $n_0$ (cm$^{-3}$). The Str\"omgren radius and value of \doh~are also provided. Entries: model number; age of cloud, $t$; dust content (0=dust-free, 1=dusty); Str\"omgren radius, $r_S$; outer radius of superbubble, $r_0$; thickness of the \hr, $r_1-r_0$; hydrogen density of the model cloud, $n(\rmn{H})$; time-scale to reach thermal and ionisation balance, $t_\rmn{eq}$; emission-line strength ratios [O$\,${\sc i}]\,$\lambda6300$/[O$\,${\sc ii}]\,$\lambda3727$ and [O$\,${\sc ii}]\,$\lambda3727$/[O$\,${\sc iii}]\,$\lambda\lambda4959,5007$; and contribution of the cluster's ejecta to the total oxygen abundance of the \hr, \doh.}
\begin{tabular*}{1\textwidth}{@{\extracolsep{\fill}}lcccccccccc}
\toprule
model &
t &
dust & 
r$_\rmn{S}$ &
r$_0$ &
$\rmn{r_1-r_0}$ & 
n$(\rmn{H})$ &
t$_\rmn{eq}$ &
$\frac{6300}{3727}$ &
$\frac{3727}{4959+5007}$ &
\doh \\
&
Myr &
&
pc & 
pc &
pc &  
cm$^{-3}$ &
Myr & 
&
&
dex \\
\cmidrule(r){1-11}
\multicolumn{11}{c}{cluster 1, n$_0=1$} \\
\cmidrule(r){1-11}
1 & 1 & 1 & 891  & 156 & 10 & 43  & 0.3 & 0.02 & 0.05 & 0.000\\
2 & 2 & 1 & 932  & 236 & 15 & 25  & 0.5 & 0.02 & 0.06 & 0.000\\
3 & 3 & 1 & 849  & 301 & 13 & 18  & 0.7 & 0.01 & 0.13 & 0.000\\
4 & 4 & 1 & 687  & 358 & 8 & 14  & 0.9 & 0.01 & 0.44 & 0.025\\
4a & 4 & 0 & 687  & 358 & 8 & 14  & 0.5 & 0.01 & 0.44 & 0.025\\
4b & 4 & 1 & 687  & 358 & 0 & 100  & 0.1 & 0.02 & 7.05 & -\\
4c & 4 & 1 & 687  & 100 & 62 & 14  & 0.9 & 0.01 & 0.06 & 0.227\\
5 & 5 & 1 & 561 & 410 & 5 & 12  & 1.1 & 0.01 & 1.91 & -\\

\cmidrule(r){1-11}
\multicolumn{11}{c}{cluster 1, n$_0=10$} \\
\cmidrule(r){1-11}
6 & 1 & 1 & 192  & 98 & 2 & 173  & 0.1 & 0.02 & 0.08 & 0.000\\
7 & 2 & 1 & 201  & 149 & 3 & 99  & 0.1 & 0.02 & 0.09 & 0.000\\
8 & 3 & 1 & 183  & 190 & - & 72  & 0.2 & 0.01 & 0.21 & -\\
9 & 4 & 1 & 148  & 226 & 1 & 57  & 0.2 & 0.01 & 0.78 & -\\
9a & 4 & 0 & 148  & 226 & 1 & 57  & 0.1 & 0.01  & 0.78 & -\\
9b & 4 & 1 & 148  & 226 & 0 & 100  & 0.1 & 0.01 & 1.66 & -\\
9c & 4 & 1 & 148  & 100 & 6 & 57  & 0.2 & 0.01 & 0.13 & 0.206\\
10 & 5 & 1 & 121  & 258 & 1 & 48  & 0.3 & 0.01 & 3.65 & -\\
\cmidrule(r){1-11}
\multicolumn{11}{c}{cluster 2, n$_0=1$} \\
\cmidrule(r){1-11}
11 & 1 & 1 & 902  & 167 & 6 & 50  & 0.2 & 0.02 & 0.07 & 0.000\\
12 & 2 & 1 & 1,144  & 254 & 16 & 29  & 0.4 & 0.02 & 0.05 & 0.000\\
13 & 3 & 1 & 1,306  & 324 & 26 & 21  & 0.6 & 0.02 & 0.04 & 0.000\\
14 & 4 & 1 & 1,390  & 385 & 36 & 16  & 0.7 & 0.02 & 0.04 & 0.008\\
14a & 4 & 0 & 1,390  & 385 & 37 & 16  & 0.4 & 0.02 & 0.04 & 0.008\\
14b & 4 & 1 & 1,390  & 385 & 1 & 100  & 0.1 & 0.02 & 0.22 & 0.002\\
14c & 4 & 1 & 1,390  & 100 & 149 & 16  & 0.7 & 0.03 & 0.01 & 0.035\\
15 & 5 & 1 & 1,430  & 440 & 43 & 14  & 0.8 & 0.02 & 0.04 & 0.018\\
\cmidrule(r){1-11}
\multicolumn{11}{c}{cluster 2, n$_0=10$} \\
\cmidrule(r){1-11}
16 & 1 & 1 & 194  & 106 & 1 & 199  & 0.1 & 0.02 & 0.11 & 0.000\\
17 & 2 & 1 & 246  & 160 & 3 & 114  & 0.1 & 0.02 & 0.07 & 0.000\\
18 & 3 & 1 & 281  & 204 & 4 & 83  & 0.2 & 0.02 & 0.06 & 0.000\\
19 & 4 & 1 & 299  & 243 & 6 & 66  & 0.2 & 0.02 & 0.06 & 0.009\\
19a & 4 & 0 & 299  & 243 & 6 & 66  & 0.1 & 0.02 & 0.06 & 0.009\\
19b & 4 & 1 & 299  & 243 & 3 & 100  & 0.1 & 0.02 & 0.08 & 0.006\\
19c & 4 & 1 & 299  & 100 & 26 & 66  & 0.2 & 0.02 & 0.01 & 0.066\\
20 & 5 & 1 & 308  & 277 & 7 & 55  & 0.2 & 0.02 & 0.06 & 0.021\\
\bottomrule
\end{tabular*}\label{tab:cl}
\end{table*}
\subsection{\hr~models}\label{res:hrs}
For each cluster and value of $n_0$, Table~\ref{tab:cl} gives input and output properties of the \hr~models, as well as some supplementary data. The entries in the table are described in the caption and are discussed in sections \ref{res:r0} to~\ref{res:doh}, except for the data corresponding to the 12 supplementary simulations at 4\,Myr, which either have no dust (models a), have $n(\rmn{H})=100$ (models b), or have $r_0=100$ pc (models c). Models a, b, and c are discussed in \S~\ref{res:extra}.
\subsubsection{Radius $r_0$}\label{res:r0}
In order to see if $r_1>r_0$ is a reasonable assumption, I computed the Str\"omgren radius, $r_\rmn{S}=\{3\,Q(\rmn{H})/[4\pi\,\alpha_A^r(\rmn{H},T)]\}^{1/3}$, of each \hr, where $\alpha_A^r(\rmn{H},T)$ is the case A radiative recombination coefficient to form \hi~in any energy level. I used the value at 10$^4$ K (characteristic of \hrs), i.e., $\alpha_A^r=4.2\times10^{-13}$ cm$^3$ s$^{-1}$ \citep{ver96}. For most models, $r_\rmn{S}>r_0$, in agreement with the assumption that the ionisation front of the \hr~is located outside of the superbubble. However, models 8, 9, 9a, 9b, and 10 have $r_\rmn{S}<r_0$. Therefore, the latter models are excluded from further analysis. 
\begin{figure*}
\centering
\includegraphics[width=0.8\textwidth]{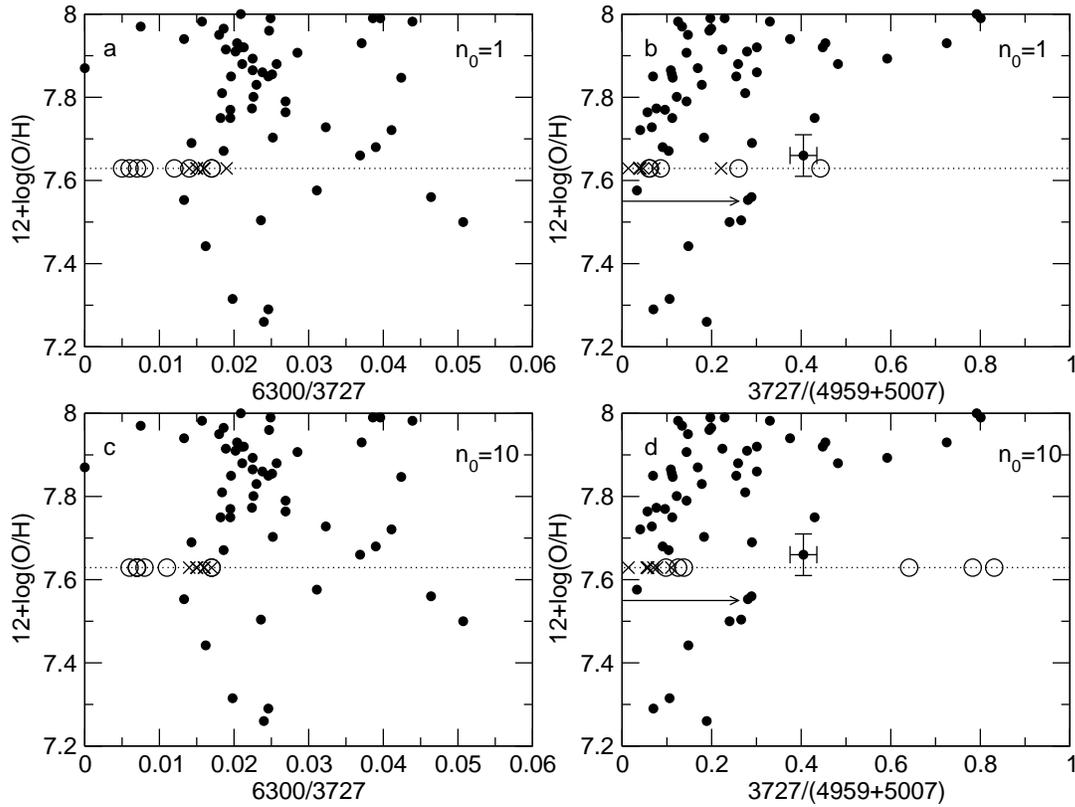}
\caption{Observed versus model oxygen emission-line strength ratios $6300/3727$ and $3727/(4959+5007)$ (including the 32 models in Table~\ref{tab:cl}). Panels: a \& b, $n_0=1$; c \& d, $n_0=10$. Symbols: filled circles, observed values collected by Nava et al. (2006); open circles, clouds ionised by cluster 1; crosses, clouds ionised by cluster 2. Vertical dotted lines: oxygen abundance of model \hrs. Arrow in panel b: direction of time for the five models 1 to 5. Arrow in panel d: direction of time for the five models 6 to 10.}\label{fig:lratios}
\end{figure*}
The radius $r_0$ increases as $t$ increases, $n_0$ decreases, and $L_{38}$ increases. Continuous star formation yields larger values of $r_0$ by a few tenths of parsecs, because $L_{38}^2>L_{38}^1$. All values of $r_0$ are within $\sim$100--400 pc, therefore, some of them are comparable to the thickness of the Galactic disk, which is $\sim$200 pc \citep{mat01}. A superbubble breaks when it reaches the halo of its host galaxy, as mentioned in \S~\ref{sec:int}. However, in the present work, I assumed that the superbubble and its associated \hr~remain within the gas-rich component of the host galaxy (which is not necessarily a disk galaxy), and that stellar ejecta is confined within radius $r_1$, from 1--5\,Myr.
\subsubsection{Hydrogen density}\label{res:nh}
The hydrogen density $n$(H) increases as $t$ decreases, $n_0$ increases, and $L_{38}$ increases. All values of $n$(H) are within $\sim$10--200 cm$^{-3}$, which is in good agreement with observed values in giant \hrs~\citep[e.g., see column 2 of table~5 in][]{nav06}. 
\subsubsection{Equilibrium time-scale}
The reason for checking the value of $t_\rmn{eq}$ is that in order to be valid, simulations must have $t_\rmn{eq}<1\,$Myr. This is because subsequent models are 1$\,$Myr apart. For all models, the longest process to reach equilibrium is heating-cooling balance. Cooling is dominated by collisional cooling due to hydrogen lines, which becomes less important as time increases and $n$(H) decreases. This is why $t_\rmn{eq}$ increases as time increases and $n$(H) decreases. The requirement that $t_\rmn{eq}$ be less than 1\,Myr rules out model 5 as valid.
\subsubsection{Oxygen emission-line ratios}\label{res:lratios}
Figure~\ref{fig:lratios} shows plots of observed versus model values of the ratios $R12=6300/3727$ and $R23=3727/(4959+5007)$. It includes the 32 models in Table~\ref{tab:cl}. The filled symbols correspond to the observational data collected by \cite{nav06}, the open symbols to clouds ionised by cluster 1, and the crosses to clouds ionised by cluster 2. All models are located on the vertical-dotted lines at $\rmn{12+log(O/H)}=7.629$, which is the input oxygen abundance of the CLOUDY models. The value of $n_0$ is indicated in each panel. The arrows shown in panels b and d indicate the direction of time for the ten models 1 to 5 and 6 to 10, respectively. For clouds ionised by cluster 1, $R23$ increases as time increases because as the most luminous stars in the cluster fade away, the amount of O$^{+2}$ decreases, whereas the amount of O$^+$ increases.

The observed values in panels a and c present no trend and cannot be used to test the validity of the models. Note that in most cases, only an upper limit is available for the observed value of [O I]$\,\lambda6300$. On the other hand, panels b and d show that the range of observed values of $R23$ increases with the oxygen abundance. This trend was used to test the validity of the models. The observational data in panels b and d includes no points with $R23$ greater than about 0.45 at $\rmn{12+log(O/H)}=7.629$. Of the remaining good models (exclude models 5, 8, 9, 9a, 9b, and 10), only 4b has $R23>0.45$. Therefore, model 4b was also excluded. Note that the point with the error bars in Fig.~\ref{fig:lratios}, is KISSR 1013 \citep{mel04}, which is a giant emission nebula with $\rmn{12+log(O/H)}=7.632$ \citep{nav06}. The horizontal error bar was computed using standard error propagation theory and was taken as the characteristic error bar of model 4, which is the open circle closest to KISSR 1013 in Fig.~\ref{fig:lratios}. Therefore, model 4 was considered to be a valid model. 
\subsubsection{Thickness of \hrs}
The radius $r_1$ is given by the point where the fraction of ionised to total hydrogen is $\rmn{H^+/H=0.9}$. The \hrs~are thin shells with thicknesses ($r_1-r_0$) equal to a few per cent of $r_0$. The \hrs~ionised by cluster 1 become thinner as the ionising stars fade away, while those ionised by cluster 2 become thicker, as the value of $Q$(H) increases.
\subsubsection{Self-enrichment}\label{res:doh}
The values of $\Delta n(\rmn{H})$ and $\Delta n(\rmn{O})$ were computed from Eqn. (\ref{eqn:deln}), using the values in columns 6 and 7 of Table~\ref{tab:feedback}, respectively. For all models, $\Delta n(\rmn{H})\le5\times10^{-3}$ and $n_i(\rmn{H})=n(\rmn{H})>10$. Therefore, Eqn. (\ref{eqn:deln}) can be simplified to
\begin{equation}
\left(\frac{\rmn{O}}{\rmn{H}}\right)_f\approx\left(\frac{\rmn{O}}{\rmn{H}}\right)_i+\frac{\Delta n(\rmn{O})}{n_i(\rmn{H})},
\end{equation}
which yields
\begin{equation}
\Delta\rmn{log(O/H)}=\rmn{log}\left[\left(\frac{\rmn{O}}{\rmn{H}}\right)_i+\frac{\Delta n(\rmn{O})}{n(\rmn{H})}\right]+4.371,
\end{equation}

Table~\ref{tab:feedback} shows that $M$(O) is only significant at 4 and 5\,Myr. The fraction of $M$(O) used to pollute the \hrs, is given by $V_1/(V_0+V_1)=1-r_0^3/r_1^3$. It is less than 0.25, for valid models, and excluding models c, which have an arbitrary value of $r_0=100$ pc. Of the models corresponding to 4 or 5\,Myr, models 4, 14, 15, 19, and 20 are good models. Of the latter five models, model 4, which corresponds to the instantaneous SFL, has the maximum value of \doh, which is 0.025 dex. This value is within the range of observational uncertainties in O/H of metal-poor \hrs~\citep[see column 3 of table 6 in][]{nav06}. In addition, recall that the value of $r_0$ is underestimated at 4\,Myr (see \S~\ref{res:mech}), therefore, \doh~may be even lower.
\subsection{Supplementary models}\label{res:extra}
I proceed to discuss the 12 supplementary models, a, b, and c at 4\,Myr. The comparisons of models 4 with 4a, 14 with 14a, and 19 with 19a, show that dust-free clouds (models a) are indistinguishable from dusty models. This is due to the low metallicity of the clouds. The comparisons of models 14 with 14b and 19 with 19b, show that increasing the value of $n(\rm{H})$ by a few $\times10$\,cm$^{-3}$ decreases the value of \doh~by a few $\times10^{-3}$\,dex, which is a negligible amount. Finally, comparisons of models 4 with 4c, 14 with 14c, and 19 with 19c, show that reducing $r_0$ by a few $\times100$\,pc, significantly increases \doh. In particular, the value of \doh~for model 4 is larger than observed oxygen abundance fluctuations in \hrs~of higher metallicities, such as those found in M101 \citep[equal to 0.1--0.2\,dex][]{ken96} and NGC 4214 \citep[equal to 0.1\,dex][]{kob96}. Note that Eqn. (\ref{eqn:r0}) for $r_0$ is based on a model that neglects the effect of gravity. If gravity was included, this would reduce the value of $r_0$, and increase the value of \doh.
\section{Summary and Conclusion}\label{sec:con}
I used version 5.1 of the evolutionary synthesis code STARBURST99 \citep{lei99,vaz05} for predicting the radiative, mechanical, and chemical feedbacks of young ($t\le5\,$ Myr) metal-poor ($Z=0.001$) stellar populations, resulting from either a single burst of star formation with a total stellar birth mass of $10^6\,\rmn{M}_\odot$ (cluster 1), or from continuous star formation at a rate of 1\ms$\,$yr$^{-1}$ (cluster 2). I adopted a Salpeter IMF from $0.1-120\,\rmn{M}_\odot$, SNII and black hole cut-off masses of 8 and 120\ms, respectively, the non-rotating evolution tracks with standard mass loss of \cite{sch92} for stars with masses $M<12\,$M$_\odot$, and the non-rotating evolution tracks with high mass loss of \cite{mey94} for stars with $M\ge12\,$M$_\odot$, and individual stellar spectra from \cite{lej97} for stars with plane-parallel atmospheres, \cite{pau01} for O stars, and \cite{hil98} for other stars with strong winds.

The radiative and mechanical cluster feedbacks predicted by STARBURST99 served as input to independent one-dimensional constant-density steady-state simulations of the emission-line spectra of metal-poor ($Z=0.001$) \hrs, at the discrete ages 1, 2, 3, 4 and 5\,Myr. The simulations were computed with version 07.02 of the plasma/molecular photoionisation code CLOUDY \cite{fer98}. My model cloud is a spherically symmetric shell of ISM, located at radius $r_0$, where $r_0$ is the outer radius of the adiabatically expanding superbubble of \cite{mac88}. I let the volume of the superbubble be $V_0$ and that of the \hr~be $V_1$. In order to explore why giant \hrs~do not show significant metal enhancements attributable to contamination with the ejecta from embedded massive stars, I modeled their self-enrichment in oxygen ejected by WR stars. This was done by uniformly and instantaneously enriching the volume $V_0+V_1$ of valid models, with stellar ejecta predicted by STARBURST99. 

I defined a valid \hr~model as one that satisfies the following three conditions. First, the Str\"omgren radius of the \hr~must be larger than $r_0$, since the \hr~was placed at $r_0$. Second, the time-scale to reach thermal and ionisation balance must be $t_\rmn{eq}<1$\,Myr, since the models are 1\,Myr apart. Third, the value of the [O\,{\sc ii}]/[O\,{\sc iii}] emission-line strength ratio, $R23=3727/(4959+5007)$, must be in agreement with observed values in metal-poor giant \hrs.

I let the contribution of massive star ejecta to the total oxygen abundance of the enriched \hr~be \doh. The effect on \doh~of changing the dust content of the cloud is negligible, due to the low metallicity of the models. The effect on \doh~of changing the value of $n$(H) by a few $\times10$\,cm$^{-3}$ is also negligible. On the other hand, reducing the value of $r_0$ by a few $\times100$\,pc, significantly increases the value of \doh. It would be interesting to include the effect of gravity in the model of the expansion of a superbubble, since it would reduce the value of $r_0$ and increase the value of \doh. My main result is that in metal-poor \hrs, self-enrichment in oxygen from WR stars is unlikely to be the cause of observed variations in O/H, larger than 0.025\,dex, over spatial-scales of a few $10^2\,$pc. This is because for objects similar to those modeled in this paper, self-enrichment due to the winds of Wolf-Rayet (WR) stars results in a maximum oxygen abundance enhancement \doh\,=\,0.025\,dex. The latter enhancement is produced by cluster 1 at 4\,Myr, and it is within the range of uncertainties in the O/H abundances of metal-poor giant \hrs~\citep[e.g.][]{nav06}.  

Prior to computing simulations of \hr~self-enrichment at ages $t>5$\,Myr, the following should be considered. First, if the ionising cluster is like cluster 1, then as time increases, the \hr~becomes fainter and harder to observe, but even worse, for $t>5$\,Myr, $t_\rmn{eq}>1$\,Myr, and also, the value of $R23$ becomes larger than observed values reported in the literature. In addition, at times $t>5$\,Myr, and for the two star formation laws under consideration, the stellar ejecta may not be confined within radius $r_0$ anymore, since the superbubble may have broke open into the halo of the host galaxy. Finally, at times $t>5$\,Myr, it is necessary to determine if the SN mechanical energy transformed into ionising diffuse radiation needs to be included as an additional source of ionisation (e.g., see \citealt{cho08}).  

In the future, it would be interesting to study \hr~self-enrichment with time-dependent CLOUDY models, and also, to include in the study, elements such as He, C and N. Also, in order to complete this study, one should analyze the effects of changing the mass and space distribution of the stars. In particular, one should keep in mind the work of \cite{erc07}, who computed using six different methods the oxygen abundances that would be observed for spherical HIIRs of constant density, $n(\rmn{H})=100$ cm$^{-3}$, radius, $r_0=9$ pc, metallicities $Z=0.020$ and $Z=0.001$, and ionised by stars with different spatial distributions, with a fixed total H-ionising photon emission rate, $Q_\rmn{H}=3.8\times10^{50}$ s$^{-1}$, such that half of the power is provided by stars with masses of $M=37\,\rmn{M}_\odot$, and the other half by stars with masses of $M=56\,\rmn{M}_\odot$. \cite{erc07} found i) that the typical difference in log(O/H) between the limiting cases where the stars occupy the same point in space, as in the present work, or are randomly distributed within the spherical cavity of radius $r_0=9$ pc, ranges from $0.1-0.3$ dex, and ii) that the value of log(O/H) that would be measured, would be smaller for the point source scenario, for five of the abundance-determination methods adopted.
\section*{Acknowledgments}
This research is supported by NSF grant AST 03-07118 to the University of Oklahoma. I am thankful to L. van Zee who triggered my interest in this topic. I am also thankful to R.B.C Henry, M. Cervi\~no, D. Schaerer, M.-M. Mac Low, E. Skillman, Y-H Chu, C. Leitherer, and the referee of this paper, for their invaluable help with this work.
\bibliography{bib}
\appendix
\section*{Appendix A: superbubble model}\label{a:sb}
As mentioned in the introduction, massive star winds and SNe sweep up and compress the ambient ISM into shells. If the mechanical energy producing the shell is provided by one star, then the shell is called a bubble, and if it is provided by several stars, then it is called a superbubble. The size of a single star bubble is a few $\times10$ pc, whereas that of a superbubble is a few $\times10^2$ pc \citep{chu04}.

\cite{wea77}, henceforth WMC77, have described the structure and evolution of an idealized bubble, which expands through a uniform ISM, driven by a spherically symmetric wind with a constant mechanical luminosity. Starting from the inside, the three distinct regions of this bubble, r1-r3, and the boundaries of each region, b1-b3, are: (r1) a freely expanding wind, (b1) a wind termination shock, (r2) a shocked wind region, (b2) a contact discontinuity region, (r3) a shocked ISM region, and (b3) a forward shock propagating into the ambient ISM (see fig. 1 in WMC77). In addition, the bubble's evolution can be divided into four stages, s1-s4: (s1) free expansion of the wind for the first few hundred years, (s2) adiabatic expansion of the bubble for the next few thousand years (since the expansion is rapid, radiative losses do not have time to affect any part of the system), (s3) snowplow phase for most of the stellar lifetime (this phase starts when radiative losses cause the ISM swept up by the bubble to cool and collapse into a thin shell, and during this phase, the shocked stellar wind still conserves energy), and (s4) time when radiative losses also affect the dynamics of the shocked wind region.

\cite{mac88}, henceforth MM88, extended the work of WMC77 to the multiple-star case, by considering the superbubble blown by the winds and SNeII of stars in an OB association. In the model of MM88, the superbubble also expands through a homogeneous ISM, since according to the authors, the interaction of the SN ejecta with the pre-existing wind bubble does not affect the later dynamics of the superbubble. In addition, the superbubble's structure is similar to that of the bubble described in WMC77, since the stellar ejecta is assumed to be thermalized and confined within the superbubble. \cite{mac88} concentrate on the evolution of the superbubble subsequent to the collapse of the swept up ISM into a thin cold shell, as in stage (s3) mentioned above. For our parameters, this collapse occurs a few $\sim10^4$ yr after the beginning of star formation, as given by eqn. (8) in MM88. Throughout this paper, I considered that the superbubble is in stage (s3). As in WMC77, the contact discontinuity region  of the superbubble is dominated by thermal conduction from the hot interior to the cold shell, and the flux of conductive energy is balanced by the flux of mechanical energy associated with evaporation of shell gas, into the interior. Accordingly, the superbubble is filled with gas which is a combination of stellar ejecta and ISM. 
\section*{Appendix B: Absorption within $V_0$}\label{a:sink}
For the purpose of radiative transfer, the volume of radius $r_0$ ($V_0$) was treated as a cavity empty of gas. Here I show that the number of H-ionising starlight photons, $Q(H)$, is not greatly reduced by absorptions within $V_0$.

First, I note that H is the dominant absorber of ionising radiation within $V_0$. As discussed in \citet[][\S~2.4]{ost06}, in the low-density limit, which holds within $V_0$, absorptions by He do not greatly reduce $Q(H)$. In addition, due to the low metallicity of the models, absorptions of ionising photons by metals do not greatly reduce $Q(H)$ either. Second, I assume that each hydrogen ionisation by a starlight photon is exactly balanced by a hydrogen recombination. Therefore, showing that $Q(H)$ is not greatly reduced within $V_0$, is equivalent to showing that the number of hydrogen recombinations within $V_0$, $A_0(H)$, is much less than $Q(H)$. Third, I use the approximations: $n\approx n(\rmn{H})+n(\rmn{He})\approx1.1\,n(\rmn{H})$, $n_e\approx n_p +n(\rmn{He^+})$, $n_p \approx n/1.1$, and $n(\rmn{He^+})\approx 0.1\,n$, where $n$, $n_e$, $n_p$, and $n(\rmn{He^+})$ are the atomic, electron, proton and He$^+$ number densities, respectively. These approximations yield: $n_en_p\approx0.92n$. Therefore, the value of $A_0(\rmn{H})$ is given by:
\begin{equation}
A_0(\rmn{H})=4\pi~\int_0^{r_4} \alpha_A^r(\rmn{H},T)\,(0.92\,n^2)\,r^2\,dr,\label{eqn:Arh}
\end{equation}
where $r_4$ is the radius where the superbubble's temperature has dropped to $10^4\,$K (which is the typical equilibrium temperature of an \hr), and $\alpha_A^r(\rmn{H},T)$ (cm$^3$ s$^{-1}$) is the case A radiative recombination coefficient to form \hi~in any energy level. For $\alpha_A^r$, I used the parametrization of \cite{ver96}. For $n$ and $T$ I used Eqns. (\ref{eqn:t}) and (\ref{eqn:n}), respectively. I found log$\,A_0(\rmn{H})<46.94\,$s$^{-1}$ for all my models. On the other hand, log$\,Q(\rmn{H})>51.95\,$s$^{-1}$ for all star clusters and times considered (see Table~\ref{tab:feedback}). Since $A_0(\rmn{H})\ll Q(\rmn{H})$, the gas within $V_0$ does not significantly absorb the ionising radiation from the stars. 
 
Note that Eqn. (\ref{eq:n}) gives a lower limit to the density inside the superbubble, since it assumes that the interior is dominated by thermal conduction, whereas \cite{dea02} strongly suggest that turbulent mixing dominates at the contact discontinuity region, i.e., they suggest that in reality, the density is higher and less homogeneous. However, the process of turbulent mixing is still under investigation and for now, I adopted Eqn. (\ref{eq:n}).
\label{lastpage}
\end{document}